\documentclass[a4paper,12pt]{article}

\usepackage{amsmath}
\usepackage{amsfonts}
\usepackage{cite}
\usepackage{float}
\usepackage[margin=27mm]{geometry}
\usepackage{listings}
\usepackage{url}
\usepackage{xcolor}
\usepackage{hyperref}

\hypersetup{
  colorlinks=true,
  linkcolor=blue,
  citecolor=blue,
  urlcolor=black
}

\definecolor{background}{gray}{0.97}
\definecolor{frame}{gray}{0.80}
\definecolor{keyword}{rgb}{0,0.6,0}
\newcommand{\basisgen}{\texttt{BasisGen}}
\newcommand{\code}{\texttt}
\newcommand{\Sym}{\operatorname{Sym}}
\newcommand{\antiSym}{\operatorname{\Lambda}}

\begin{document}

\begin{center}
  \textbf{\Large \basisgen: automatic generation of operator bases}

  \vspace{8mm}
  
  \textbf{J. C. Criado\footnote{jccriadoalamo@ugr.es}}

  \vspace{5mm}
  
  \emph{
    CAFPE and Departamento de F\'{\i}sica Te\'orica y del Cosmos,\\
    Universidad de Granada, Campus de Fuentenueva, E-18071 Granada, Spain
  }
\end{center}

\vspace{10mm}

\begin{abstract}
  \basisgen\ is a Python package for the automatic generation of bases of
  operators in effective field theories. It accepts any semisimple symmetry
  group and fields in any of its finite dimensional irreducible
  representations. It takes into account integration by parts redundancy and,
  optionally, the use of equations of motion. The implementation is based in
  well-known methods to generate and decompose representations using roots and
  weights, which allow for fast calculations, even with large numbers of fields
  and high-dimensional operators. \basisgen\ can also be used to do some
  representation-theoretic operations, such as finding the weight system of an
  irreducible representation from its highest weight or decomposing a tensor
  product of representations.
\end{abstract}

\vfill

\newpage

\section{Introduction}%
\label{sec:introduction}

Effective field theory is a widely used framework for parameterizing the physics
of systems whose degrees of freedom and symmetries are known. An effective
Lagrangian is a linear combination of all local operators that can constructed
with the fields in the theory, with the restriction that they are invariant
under the action of the symmetry group. Usually, there is some other constraint
that reduces the number of possibilities to a finite one, such as imposing a
maximum canonical dimension. In this context, it is often convenient to obtain a
complete set of independent operators, which is called a basis. \basisgen\
automatizes this task.

The input data needed for this calculation are the symmetry group $G$ of the
theory and the representation of $G$ corresponding to each field. Once they are
specified, one can obtain, for every monomial in the fields, the number of
independent ways of forming an invariant under the action of $G$ out of it. It
must also be taken into account that total derivative terms can be added to the
Lagrangian without changing the physics (except for effects of surface terms in
the action). This means that some operators with derivatives can be rewritten in
terms of others. Moreover, at each order in the effective Lagrangian, the
addition of an operator proportional to the equations of motion does not change
the S matrix up to higher order effects~\cite{%
  Politzer:1980me,GrosseKnetter:1993td,Arzt:1993gz,Wudka:1994ny,Criado:2018sdb%
}. It follows that the equations of motion can be used, for example, to obtain a
basis in which all the operators proportional to the functional derivative of
the kinetic term have been removed~\cite{%
  Georgi:1991ch,Grzadkowski:2003tf,Fox:2007in,AguilarSaavedra:2008zc,%
  AguilarSaavedra:2009mx%
}. For the Standard Model Effective Field Theory (SMEFT) (see
ref.~\cite{Brivio:2017vri} for a review), several bases and (incomplete) sets of
independent operators have been computed taking all these facts into
account~\cite{%
  Hagiwara:1993ck, Giudice:2007fh, Grzadkowski:2010es, Elias-Miro:2013eta%
}. Computer tools can be used to translate from one basis to another~\cite{%
  Falkowski:2015wza, Criado:2017khh, Aebischer:2018bkb, Gripaios:2018zrz%
}.

In the last few years, many developments have been made in the automatization of
the generation of operator bases. Hilbert series methods provide an elegant way
to compute invariants~\cite{%
  Lehman:2015via,Henning:2015daa,Lehman:2015coa,Henning:2015alf,Henning:2017fpj%
}. They can be directly implemented in a computer system with symbolic
capabilities, as done for the SMEFT case in the auxiliary \code{Mathematica}
notebook of ref.~\cite{Henning:2015alf}. One possible drawback of this approach,
when used in computer code, is its performance, as an overhead due to the
symbolic nature of the calculations might be introduced.  The program
\code{DEFT}~\cite{Gripaios:2018zrz}, written in \code{Python}, uses a different
approach to check and generate bases of operators for theories with a symmetry
group given by a product of unitary groups.

\basisgen\ uses yet another approach, which is valid for any semisimple symmetry
group and avoids the need for symbolic calculations. The algorithms that it uses
to deal with representations of semisimple Lie algebras are the classical ones,
based weight vectors. They are reviewed, for example, in
ref.~\cite{Slansky:1981yr}, and implemented in several computer packages with
different purposes~~\cite{%
  Leeuwen:1992, Candiello:1994zj, Fischbacher:2002fr, Horst:2010qj,
  Nazarov:2011mv, Feger:2012bs%
}. To remove integration by parts redundancy, an adaptation of the method in
ref.~\cite{Henning:2017fpj} is used. \basisgen\ is $\sim 150$ times faster than
the implementation in the auxiliary notebook of ref.~\cite{Henning:2015alf}. For
example, \basisgen\ takes 3 seconds to compute the 84 dimension-6 operators of
the 1-generation SMEFT (in a laptop with a 2,6 GHz Intel Core i5 processor),
while the notebook of ref.~\cite{Henning:2015alf} takes $~7$ minutes. According
to ref.~\cite{Gripaios:2018zrz}, \code{DEFT} also takes minutes for the same
calculation.

For computations with effective field theories, \basisgen\ assumes 4-dimensional
Lorentz invariance. In addition, an internal symmetry group must be specified.
This is, in general, the product of the global symmetry group and the gauge
group. Derivatives are assumed to be gauge-covariant derivatives, so that the
derivative of any field has the same representation under the internal symmetry
group as the field itself. The gauge field strengths to be included in a
calculation should be provided by the user. The fields must belong to linear
irreducible representations of both the Lorentz group and the internal symmetry
group. Finally, it is required that a power counting based in canonical
dimensions can be used.

In this context, \basisgen\ generates bases of invariant operators. Sets of all
possible covariant operators, with their corresponding irreducible
representations (irreps), can also be computed. The basic
representation-theoretic functionalities needed for these calculations are:
obtaining weight systems of irreps and decomposing their tensor products. An
interface for their direct use is provided.

\basisgen\ can be installed using \code{pip} by doing: \code{pip install
  basisgen}. It requires \code{Python} version 3.5 or higher. Its code can be
downloaded from the GitHub repository
\url{https://github.com/jccriado/basisgen}, where some examples of usage can be
found. A simple script using \basisgen\ is presented in
listing~\ref{lis:simple}. It defines an effective theory with internal symmetry
group $SU(2) \times U(1)$ for a complex scalar $SU(2)$-doublet field with charge
$1/2$. It computes a basis of operators of dimension 8 or less. The output is
presented in listing~\ref{lis:simple-output}. Each line gives the number of
independent invariant operators that can be constructed with each field content.

\begin{samepage}
  The rest of this article is divided in two sections (apart from the
  conclusions). They describe \basisgen's implementation
  (section~\ref{sec:implementation}) and interface
  (section~\ref{sec:interface}).

\end{samepage}

\begin{lstlisting}[
  caption=Simple EFT example script,
  float=t,
  label=lis:simple
  ]
from basisgen import algebra, irrep, scalar, Field, EFT

phi = Field(
    name='phi',
    lorentz_irrep=scalar,
    internal_irrep=irrep('SU2', '1'),
    charges=[1/2]
)

my_eft = EFT(algebra('SU2'), [phi, phi.conjugate])

invariants = my_eft.invariants(max_dimension=8)

print(invariants)
print("Total:", invariants.count())
\end{lstlisting}

\begin{lstlisting}[
  caption=Simple EFT example script's output,
  float=t,
  label=lis:simple-output
  ]
phi phi*: 1
(phi)^2 (phi*)^2: 1
(phi)^2 (phi*)^2 D^2: 2
(phi)^2 (phi*)^2 D^4: 3
(phi)^3 (phi*)^3: 1
(phi)^3 (phi*)^3 D^2: 2
(phi)^4 (phi*)^4: 1
Total: 11
\end{lstlisting}

\section{Implementation}%
\label{sec:implementation}

\subsection{Basic operations with representations}%
\label{sec:representations}

In this section, the methods implemented in \basisgen\ to deal representations
of semisimple Lie algebras are presented. A representation of a semisimple
algebra is just a tensor product of representations of the algebra's simple
ideals. Using this fact, \basisgen\ decomposes calculations with semisimple
algebras into smaller ones with simple algebras. The basic operations with
representations of simple algebras are: the generation of the weight system of
an irrep from its highest weight and the decomposition of a reducible
representation into a direct sum of irreps. They are both implemented using
well-known methods (see refs.~\cite{%
  Slansky:1981yr, Leeuwen:1992, Candiello:1994zj, Fischbacher:2002fr,
  Horst:2010qj, Nazarov:2011mv, Feger:2012bs%
}), which are summarized here, for completeness.

In the Dynkin basis, which we use in what follows, all weights are tuples of
integers. Thus, the operations done here involve only addition and
multiplication of integer numbers. Each irrep of a simple algebra is uniquely
characterized by its highest weight $\Lambda$, which is a tuple
$(a_1 \ldots a_n)$ of non-negative integers. Every such tuple is the highest
weight of one irrep. The complete weight system of an irrep may be obtained from
its highest weight by the following procedure:
\begin{enumerate}
\item Set $W = \{\}$ and $W_{\text{new}} = \{\Lambda\}$.
\item Choose some $\lambda \in W_{\text{new}}$.
\item For each positive component $\lambda_i > 0$, select the $i$th row $\alpha$
  of the Cartan matrix. Append to $W_{\text{new}}$ all weights of the form
  $\lambda - k \alpha$, with $0 < k \leq \lambda_i$.
\item Remove $\lambda$ from $W_{\text{new}}$. Append it to $W$.
\item If $W_{\text{new}}$ is empty, terminate. Otherwise, go to step 2.
\end{enumerate}
This produces the set $W$ of all weights. The multiplicity $n_\lambda$ of each
weight $\lambda$ can then be obtained recursively using the Freudenthal formula:
\begin{equation}
  n_\lambda
  =
  \frac{
    2
    \sum_{\alpha}
    \sum_{k > 0}
    n_{\lambda + k \alpha}
    \left(
      \lambda + k \alpha,
      \alpha
    \right)
  }{
    \left(
      \Lambda + \delta,
      \Lambda + \delta
    \right)
    - \left(
      \lambda + \delta,
      \lambda + \delta
    \right)
  },
\end{equation}
where $\delta = (11 \ldots 1)$ and the summation for $\alpha$ runs over all
positive roots.

The algorithm for the decomposition of a reducible representation as a direct
sum of irreps is straightforward: from the collection of weights of the
representation in question, find the highest and remove from the collection all
the weights in the corresponding irrep. Repeat until the collection is
empty. Then, the successive highest weights that were found in the process are
the highest weights of the irreps in the decomposition. A direct application of
this functionality is to decompose the tensor product of irreps. Let $W_1$ and
$W_2$ be the weight systems of two representations $R_1$ and $R_2$. The weight
system $W$ of $R_1 \otimes R_2$ is the collection of all $\lambda_1 + \lambda_2$
for $(\lambda_1, \lambda_2) \in W_1 \times W_2$. Once $W$ is constructed, it can
be decomposed using the general decomposition algorithm.

In some cases, the symmetric or anti-symmetric tensor power of some
representation is needed. If $W = {\{\lambda_i\}}_{i \in \{1, \ldots, n\}}$ is
the weight system of some representation $R$, the weight system of the symmetric
tensor power $\Sym^k(R)$ is the collection of weights computed as
$\lambda_1 + \cdots + \lambda_k$ for every $k$-tuple
$(\lambda_{i_1}, \ldots, \lambda_{i_k})$ where $i_1 \leq \cdots \leq i_k$. The
weight system of the anti-symmetric power $\antiSym^k(R)$ is constructed in a
similar way, but using all $k$-tuples $(\lambda_{i_1}, \ldots, \lambda_{i_k})$
with $i_1 < \cdots < i_k$ instead.

\subsection{Constructing invariants in effective theories}
\label{sec:derivatives}

\basisgen\ can do calculations for 4-dimensional Lorentz-invariant effective
field theories whose internal symmetry group is of the form $G \times {U(1)}^n$,
where $G$ is semisimple. An effective theory is specified when the following
data are provided:
\begin{itemize}
\item The semisimple Lie algebra $\mathfrak{g}$ of $G$.
\item A collection of fields $\phi_1, \ldots, \phi_m$. Each $\phi_i$ must be
  equipped with:
  \begin{itemize}
  \item An irrep $R^{(i)}_{\text{Lorentz}}$ of the Lorentz algebra
    $\mathfrak{su}_2 \oplus \mathfrak{su}_2$.
  \item An irrep $R^{(i)}_{\text{internal}}$ of $\mathfrak{g}$.
  \item A tuple $\left(c^{(i)}_1, \ldots, c^{(i)}_n\right)$ of charges under the
    $U(1)$ factors.
  \item The statistics $S_i$. Either boson or fermion.
  \item A positive real number $d_i$, specifying the canonical dimension of the
    field.
  \end{itemize}
\end{itemize}
It is assumed that a power counting based in canonical dimensions of the fields,
with derivatives having dimension 1, can be applied. This is used to reduce the
number of possible operators to a finite one.

The main functionality of \basisgen\ is to compute the number of independent
invariant operators, constructed with the fields $\phi_i$ and their (covariant)
derivatives, and having dimension less than or equal to some fixed
$d_{\text{max}}$. To do this, first, all the possible operator field contents
are found. The field content for some operator is identified by a tuple
$\mathcal{C} = (e_1, \ldots, e_m)$, representing the exponents of each field in
the operator: $\mathcal{O} \sim {(\phi_1)}^{e_1} \cdots {(\phi_m)}^{e_m}$. For
each $\mathcal{C}$, the following (possible reducible) representation is
computed:
\begin{equation}
  \operatorname{Rep}(\mathcal{C}) =
  T_1^{e_1}(R^{(1)})
  \otimes
  \cdots
  \otimes
  T_m^{e_m}(R^{(m)}),
\end{equation}
where $T_i^k(V)$ is the symmetric power $\Sym^k(V)$ if the statistics $S_i$ are
bosonic, and the anti-symmetric power $\antiSym^k(V)$ if they are fermionic.
Once $\operatorname{Rep}(\mathcal{C})$ is obtained, it is decomposed into a
direct sum of irreps. The number of independent invariant combinations of the
fields in $\mathcal{C}$ is then easily obtained as the number of singlet irreps
in the decomposition.

To take into account (covariant) derivatives, the same procedure is used, but
now including the fields $D_\mu \phi_i$, $\{D_\mu, D_\nu\} \phi_i$,
etc. Anti-symmetric combinations of derivatives are automatically discarded, as
they are equivalent to field strength tensors. Optionally, the equations of
motion of the fields can be applied. This means that, for each
$D_{\mu_1}\ldots D_{\mu_m}\phi_i$, only the totally symmetric representation is
retained (see ref.~\cite{Lehman:2015coa}). Integration by parts redundancy can
be eliminated by removing from the results all irreps obtained from the
decomposition of $D_\mu \mathcal{O}$, $\{D_\mu, D_\nu\}\mathcal{O}$, etc.

\section{Interface}%
\label{sec:interface}

\subsection{Basic objects}%
\label{sec:basic}

\begin{minipage}{\linewidth}
  The basic objects for the usage of \basisgen\ are presented here. All of them
  can be imported with:
  \begin{lstlisting}
from basisgen import (
    algebra, irrep, Field, EFT, boson, fermion,
    scalar, L_spinor, R_spinor, vector, L_tensor, R_tensor
)
  \end{lstlisting}
\end{minipage}

\subsubsection*{Functions}%
\label{sec:functions}

\begin{description}
\item[\code{algebra}] Creates a (semi)simple Lie algebra from one string
  argument. The returned object is of the class \code{SimpleAlgebra} or
  \code{SemisimpleAlgebra} from the module \code{algebra}.
  
  \emph{Examples of arguments:} \code{'A3'}, \code{'C12'}, \code{'F4'},
  \code{'SU3'}, \code{'B2 + E7'}, \code{'SU5 x SO6 x Sp10'}.

\item[\code{irrep}] Creates an irreducible representation from 2 string
  arguments: the first represents the algebra and the second the highest weight%
  \footnote{%
    The highest weights for many irreps of several groups can be found, for
    example in~\cite{Slansky:1981yr}. In particular, notice that the highest
    weight of an $SU(2)$ irrep is its dimension minus one.%
  }. The returned object is of the class \code{representations.Irrep}.
  
  \emph{Example:} \code{irrep('SU4 x Sp7', '1 0 1 0 2 1')}. 
  
  The weight system of the a \code{representations.Irrep} object can be shown by
  calling its \code{weights\_view} method. Irreps with the same algebra can be
  multiplied to get the decomposition of their tensor product. Any two irreps
  can be added to give an irrep of the direct sum of their algebras.

  \emph{Examples, showing the weights of the octet irrep of $SU(3)$ (which has
    highest weight $(1 1)$) and the decomposition of the product of a triplet
    $(1 0)$ and an anti-triplet $(0 1)$ as an octet plus a singlet:}
  \begin{lstlisting}
>>> irrep('SU3', '1 1').weights_view()
    (1 1)
(2 -1) (-1 2)
 (0 0) (0 0)
(1 -2) (-2 1)
   (-1 -1)
>>> irrep('SU3', '1 0') * irrep('SU3', '0 1')
[1 1] + [0 0]
  \end{lstlisting}
\end{description}

\subsubsection*{Classes}%
\label{sec:classes}

\begin{description}
\item[\code{Field}] Has an attribute \code{conjugate}, the conjugate
  field. Constructor arguments:

  {\hspace{-8mm}
    \begin{tabular}{lll}
      Name &
      Description &
      Default \\
      \hline

      \code{name} &
      String identifier &
      \\

      \code{lorentz\_irrep} &
      Lorentz group irrep &
      \\

      \code{internal\_irrep} &
      Irrep of the internal (semisimple) symmetry group &
      \\

      \code{charges} &
      Charges under an arbitrary number of $U(1)$ factors &
      \code{[]} \\

      \code{statistics} &
      Either \code{boson} or \code{fermion} &
      \code{boson} \\

      \code{dimension} &
      Canonical dimension of the field &
      \code{1} \\

      \code{number\_of\_flavors} &
      Number of different copies of the same field &
      \code{1}
    \end{tabular}
  }

\item[\code{EFT}] Constructor arguments:
  \begin{description}
  \item[\code{internal\_algebra}] The semisimple Lie algebra of the internal
    symmetry group.
  \item[\code{fields}] A list of \code{Field} objects representing the field
    content of the theory.
  \end{description}
  
  Methods:
  \begin{description}
  \item[\code{invariants}] Returns a basis of operators, encapsulated in an
    \code{EFT.Invariants} object. These can be directly printed (implement
    \code{\_\_str\_\_}). They have a method \code{count} to calculate the total
    number of operators in the basis, and a method \code{show\_by\_classes},
    which returns a simplified string representation of the basis, provided a
    dictionary whose keys are the fields and values are strings representing
    classes of fields.
  \item[\code{covariants}] Returns a collection
    of all operators with all possible irreps, in the form of a
    \code{EFT.Covariants} instance. Its only purpose is to hold the information
    until it is printed (implements \code{\_\_str\_\_}).
  \end{description}
  Both receive the same arguments: \code{max\_dimension}, the maximum dimension
  of the operators computed; \code{use\_eom} (default: \code{True}) a boolean to
  specify whether the equations of motion should be used;
  \code{ignore\_lower\_dimension} (default: \code{False}), a boolean to specify
  whether operators with dimension less than \code{max\_dimension} should be
  included in the results; and \code{verbose} (default: \code{False}), a boolean
  enabling/disabling messages about the progress of the calculations.
\end{description}

\subsubsection*{Other}%
\label{sec:other}

The following irreps of the Lorentz group have been defined, for ease of use:
\code{scalar}, \code{L\_spinor}, \code{R\_spinor}, \code{vector},
\code{L\_tensor}, \code{R\_tensor}. \code{L\_spinor} and \code{R\_spinor}
correspond to left and right Weyl spinors, respectively. \code{L\_tensor} and
\code{R\_tensor} correspond to the left and right parts of an antisymmetric
tensor with two indices.

The statistics of a field can be specified by using the variables \code{boson}
and \code{fermion}, which are set to the values \code{BOSON} and \code{FERMION}
of the enum class \code{Statistics} from the module \code{statistics}.

\subsection{The \code{smeft} module}
\label{sec:smeft}

The \code{smeft} module contains the definitions of all the Standard Model
fields:
\begin{itemize}
\item The Higgs doublet \code{phi} and its conjugate \code{phic}.
\item The left and right parts \code{GL} and \code{GR} of the $SU(3)$ field
  strength.
\item The left and right parts \code{WL} and \code{WR} of the $SU(2)$ field
  strength
\item The left and right parts \code{BL} and \code{BR} of the $U(1)$ field
  strength.
\item The quark doublet \code{Q} and its conjugate \code{Qc}.
\item The lepton doublet \code{L} and its conjugate \code{Lc}.
\item The up-type quark singlet \code{u} and its conjugate \code{uc}.
\item The down-type quark singlet \code{d} and its conjugate \code{dc}.
\item The electron singlet \code{e} and its conjugate \code{ec}.
\end{itemize}
The bosons are objects of the \code{Field} class. The fermions are functions
that take the number of generations and return a \code{Field}. Similarly, the
function \code{smeft} takes the number of fermion flavors and returns an
\code{EFT} object representing the SMEFT. The algebra
$\mathfrak{su}_3 \oplus \mathfrak{su}_2$ is named
\code{sm\_internal\_algebra}. A dictionary named \code{sm\_field\_classes} is
included, to simplify the presentation of the results by passing it as an
argument to the method \code{show\_by\_classes} of an \code{EFT.Invariants}
object.

Listing~\ref{lis:smeft} contains an example script for the computation of bases
of arbitrary dimension (passed as an argument to the script) for the
1-generation SMEFT.  It gives 84 operators for dimension 6 (in about 3 seconds
in a personal computer with a 2,6 GHz Intel Core i5 processor) and 993 operators
for dimension 8 (in around 40 seconds in the same computer).
\begin{lstlisting}[
  caption=SMEFT example,
  float,
  floatplacement=H,
  label=lis:smeft
  ]
from basisgen.smeft import smeft, sm_field_classes
import sys

invariants = smeft(number_of_flavors=1).invariants(
    max_dimension=int(sys.argv[1]),
    verbose=True,
    ignore_lower_dimension=True
)

print(invariants.show_by_classes(sm_field_classes(1)))
print("Number of invariants: {}".format(invariants.count()))
\end{lstlisting}

\section{Conclusions}%
\label{sec:conclusions}

\basisgen\ computes bases of operators for effective field theories in a general
setting: the internal symmetry group can be any product of a semisimple group
and an arbitrary number of $U(1)$ factors. 4-dimensional Lorentz invariance is
assumed to provide support for concrete applications, although adaptations to
other spacetime dimensions can be easily made, due to the generality of the core
functionalities.

The decision of using the equations of motion is left to the user, as it may be
convenient to work with redundant bases in some cases (see
ref.~\cite{Criado:2018sdb}). It is also possible not only to compute invariants
but to generate all covariant operators, classified by their irreps. This can be
useful, for example, to find the representation of fields that couple linearly
to an already known theory, which are often the most relevant ones for
phenomenology~\cite{%
  delAguila:2000rc, delAguila:2008pw, delAguila:2010mx, deBlas:2014mba,
  deBlas:2017xtg%
}. An interface for doing basic operations with representations of semisimple
groups is also provided.

\basisgen's speed for large numbers of fields and high-dimensional operators
makes it possible to calculate bases for the SMEFT or for other effective
theories for physics beyond the Standard Model, in times ranging from seconds
(for the dimension-8 operators in the SMEFT) to minutes (for higher-dimensional
operators or larger number of fields) in personal computers.

\section*{Acknowledgments}

The author would like to thank M. P\'erez-Victoria for useful discussions and
comments. This work has been supported by the Spanish MINECO project
\mbox{FPA2016-78220-C3-1-P} (Fondos FEDER), the Junta de Andaluc\'ia grant
FQM101 and the Spanish MECD grant FPU14.

\bibliographystyle{utphys}
\bibliography{references}{}

\end{document}